# Identity and divergence of protein domain architectures after the Yeast Whole Genome Duplication event

D. Fusco [1,‡], L. Grassi [2,‡], A. L. Sellerio [1,‡], D. Corà [2,3], B. Bassetti[1,4], M. Caselle [2,3], M. Cosentino Lagomarsino [1,4*]

[1] *Università degli Studi di Milano, Dip. Fisica. Via Celoria 16, 20133 Milano, Italy.*

[2] *Università degli Studi di Torino, Dip. Fisica Teorica, Via Giuria 1, 10125 Torino, Italy.*

[3] *I.N.F.N. Torino, Via Giuria 1, 10125 Torino, Italy*

[4] *I.N.F.N. Milano, Via Celoria 16, 20133 Milano, Italy*

∗ : corresponding author, email: Marco.Cosentino-Lagomarsino@unimi.it, Tel. +39 02 50317477 ;

‡ : equal contribution.

## Abstract

Analyzing the properties of duplicate genes during evolution is useful to understand the development of new cell functions. The yeast *S. cerevisiae* is a useful testing ground for this problem, because its duplicated genes with different evolutionary birth and destiny are well distinguishable. In particular, there is a clear detection for the occurrence of a Whole Genome Duplication (WGD) event in *S. cerevisiae*, and the genes derived from this event ("WGD paralogs") are known. We studied WGD and non-WGD duplicates by two parallel analysis based on structural protein domains and on Gene Ontology annotation scheme respectively. The results show that while a large number of "duplicable" structural domains is shared in local and global duplications, WGD and non-WGD paralogs tend to have different functions. The reason for this is the existence of WGD and non-WGD specific domains with largely different functions. In agreement with the recent findings of Wapinski and collaborators (Nature 449, 2007), WGD paralogs often perform "core" cell functions, such as translation and DNA replication, while local duplications associate with "peripheral" functions such as response to stress. Our results also support the fact that domain architectures are a reliable tool to detect homology, as the domains of duplicates are largely invariant with date and nature of the duplication, while their sequences and also their functions might migrate.

# Introduction

Genomes possess a high degree of redundancy in the information they encode for [1, 2, 3, 4, 5]. Considering protein-coding genes, there is strong evidence [6, 7] that this redundancy has arisen from gene duplication events. Such duplications can involve individual genes, genomic segments or whole genomes. The yeast *S. cerevisiae* has arisen from an ancient whole-genome duplication [8].

The study of gene duplications is useful for understanding the evolution of proteins. Proteins descending from a common ancestor (*homologs*) are usually identified by sequence alignment methods. However, such methods typically have two main hindrances: (**i**) not taking into account directly the protein folding, which persists on longer evolutionary time scales than protein sequence, and (**ii**) being computationally intensive. On the other hand, structure and function of proteins can be described on a coarser scale, considering the protein *domains*, modular substructures that are defined by folding [9], compact structure [10], function and evolution [11]. Several authors [12, 13, 14] proved the usefulness of structural domain assignments in identifying homology. This implies that the duplicates tend to maintain their structures. This observation raises two interesting questions. The first one is how reliable the structural homology assignment is and whether it provides insight about the evolution of duplicates. The second question is whether it is possible to use domain architecture information combined to functional annotation for the characterization of duplicates from global and local duplications at different dates.

We addressed the first question by implementing an algorithm for detecting homology via structural domain assignments and comparing the results with the ones obtained by sequence alignment methods. More specifically, the description of genes at the protein domain level requires: (**i**) the construction of a protein domain *architecture* database, containing a description of each protein, in term of the domains that form it; (**ii**) the implementation of homology criteria between the entries of the database. This method is limited by our partial knowledge of protein domains, so that the architecture data suffer from incomplete coverage. Furthermore, the choice of an homology criterion implies a trade-off between error tolerance and the rate of false-positive homologs. We studied the evolution of WGD paralogs by comparing the structural domain architecture also considering their K. waltii orthologs.

The second question arises from the fact that gene duplications drive evolutionary innovation, by providing raw material to develop new functions. In particular it is interesting to understand how the whole-genome duplication event reshape the genome in a distinct way from local duplications and how this is reflected by the domain structure of duplicates. We used our method for evaluating differences between WGD paralogs and non-WGD paralogs, and performed a parallel comparison using a Gene Ontology enrichment analysis. Both analyses converge on the conclusion that whole-genome and local duplicates tend to be functionally different. Generally, core functions are enriched for WGD paralogs, while peripheral functions are enriched for non-WGD paralogs. Since domain structures of duplicates are essentially maintained, this dichotomy can be created by two main factors, for both of which we find evidence. The first one is the difference between domains that are preferentially duplicated in global and in local duplications. The second one is the migration of subcellular localization and specificity for given biological pathways.

# Results

## Homology assignment by domain characterization

The superfamily domain coverage spans one third of the genomes we examined. According to the SUPERFAMILY database, v. 1.69 [15], for *S. cerevisiae* there is a total of 6702 sequences, 3346 (50%) of which with at least one assignment. The coverage is approximately 34% of total sequence, and 85% of domains are produced by duplication. The figures for *K. waltii* are similar: 2932 (56%) sequences on 5214 were given at least one assignment, representing the 36% of total sequence covered; 84% of domains are produced by duplication.

In order to study homology from the structural domain viewpoint, we implemented three homology criteria based on domain architectures [14]. Criterion **A** defines two proteins as homologs if their domains architectures coincide (i.e. they contain the same domains in the same order). Criterion **B** allows for multiple repetitions of the same domains. The biological hypothesis behind this criterion is that, after duplication, changes may occur to the architecture of the proteins, by mechanisms such as internal duplication (e.g. by unequal crossing over), generating architectures containing multiple repetitions of one or more ancestral domains. Finally, two proteins are identified as homologs by criterion **C** if their architectures are equal, or if one of them is an approximate repetition of the other (see Methods). Biologically, this choice is motivated by the fact that it allows to recognize the simplest events of recombination, and is more fault-tolerant to differences in structural assignments generated by lack of knowledge, i.e. gaps in domain architectures.

We compared homology classes generated by the three criteria with those defined by sequence alignment methods. This test was divided in two different steps.

First, we evaluated the fraction of homology relationships identified for the WGD (by Kellis et al [8]) and by general sequence alignment methods (Ensembl-Compara [16]) that are also identified by criteria **A**, **B**, and **C**.

The results of this analysis are shown in Table 1. These results confirm the efficiency of domain-based classifications in detecting evolutionary relatedness among proteins (as observed in [17]). Specifically, they indicate that even the most stringent homology criterion **A**, is able to find the majority of triplets (72%), pairs (67%) and *Ensembl Compara* homology classes (64%). The other criteria perform better; in particular, criterion **C** retrieves more than 90% of the information in blocks of conserved syntheny. The results indicate that this method detects every type of homology (orthology and paralogy, both general and WGD).

Secondly, we quantified the fraction of paralogs not recovered by Ensembl, for each paralogy class defined by the three homology criteria (figure 1). All three criteria define a significant fraction of classes that are not recovered by Ensembl. Notice that criteria **A** and **B** follow qualitatively similar trends and produce a small fraction of partially covered classes, while criterion **C** has a larger number of partially covered classes, essentially due to the fact that classes produced with this criterion are very large. Criterion **B** is the most efficient of the three criteria in returning *Ensembl Compara* paralogy relations. Figure 1 shows the limitations of both criterion **A** and **C**. The former, being more restrictive, builds small homology classes and consequently the probability that a whole class is not recognized by Ensembl is higher. The latter builds wide homology classes associating far away homologs. The consequence is that the classes built with criterion **C** will almost certainly contain some Ensembl homologs, as shown by the small number of classes that are not recovered. On the other hand, the same classes rarely contain Ensembl homologs *only* and consequently are rarely completely covered.

## Domain architecture evolution in WGD and non-WGD duplicates

Duplicate gene pairs must undergo an altered selective regime that leads to an asymmetry emerging at different levels, for example as an increase in the rate of protein sequence evolution. Furthermore, genes at fixation may evolve in different ways, depending on the divergence process and the nature of the duplication [18]. Among the possibilities, there is a process by which one copy maintains the original function, and thus is constrained by selection, leaving the other one free to evolve, as originally hypothesized by Ohno [19] and supported by evidence in yeast [5, 20, 21]. However, theoretical and experimental work has argued that both duplicates can evolve independently at the same rate [22, 23]. We considered the question of testing the consequences of these processes at the domain level.

We followed the evolution of WGD duplicates through their domain architectures, i.e. the ordered sequence of domains forming the proteins. The length of an architecture is the total number of domains and gaps that form it. There are three main processes that affect architecture evolution: (**i**) growth by *internal* duplication, (**ii**) sequence divergence leading to structural changes in domains and (**iii**) domain insertions. In order to quantify globally the changes in protein architectures, we introduced two scoring methods that define a quantitative notion of relatedness between architectures. The first, called "domain score" is the number of domain types shared by two proteins over the sum of all domains of both proteins taken only once. The second "architecture score" measures the longest exactly matching sequence of domains between two architectures, divided by the mean length of the two architectures (see Methods).

To test for asymmetry, we compared for each WGD triplet the two *S. cerevisiae* WGD paralogs with the respective K. waltii ortholog, detecting the best- and worst- matching paralog. This was done using the domain score and the architecture score between both paralogs and their *K. waltii* ortholog. Table 2 shows the fraction (F2) of WGD triplets in which both *S. cerevisiae* duplicates have identical domain (or architecture) scores to their WGD ortholog in *K. waltii.* Furthermore we called F1 the fraction of triplets in which only one of the two *S. cerevisiae* paralogs has domain (or architecture) score one with the corresponding *K. waltii* ortholog. Comparing proteins with the architecture score we detect 65% of F2 triplets and 16% of F1 triplets, while by using the less restrictive domain score we detect 80% of F2 triplets and 11% of F1 triplets 2. This indicates that some duplicate proteins tend to evolve without changing their domain composition but rather by changing their order. We compared these results with a null model that performed random shuffling of the empirical values of the scores between the fixed sets of ortholog pairs (thus erasing the correlation between scores of the same triplet, see Methods). Interestingly, the randomized histograms show specular trends for the distribution of the scores for the two graphs that are more enhanced than for the empirical case. These data indicate that the difference between rates of "divergence" of protein architectures of WGD paralogs compared to their ortholog in *K. waltii* is larger in randomized instances. Consequently, the domain architectures of WGD duplicates are typically more balanced than expected from the null model, rather than more asymmetric.

We extended the analysis of paralog divergence to non-WGD paralogs, taking into account the duplication date reported by Wapinski *et al.* [24]. Measuring the average domain and architecture scores as a function of duplication age, and their standard deviations on the age sets, we find that the domain score is roughly constant and very close to one (figure 2), indicating that even ancient paralogs maintain similar domain composition. The more stringent architecture score shows a similar trend, with a more marked decrease for pre-WGD paralogs. Note that the proximity to one of the domain score implies that the same score for single-copy orthologs cannot be much higher, and thus that the observed accelerated evolution of paralogs [20] should not be seen at the domain level. In order to test this directly, we have considered the distribution of domain and architecture score for single-copy *S. cerevisiae* genes versus their *K. waltii* orthologs and we have compared this result with domain and architecture score of double-copy *S. cerevisiae* genes (WGD duplicates) versus their *K. waltii* orthologs. The two histograms perfectly overlap for both architecture and domain score (figure S1).

## Functional divergence and duplication age

In order to gain more insight into the divergence of duplicates at the domain level, we evaluated how the same duplicate proteins tend to diverge in their function. Specifically, we calculated the Gene Ontology (GO) term similarity between paralogs for each of the GO branches ("molecular function", "biological process" and "cellular component") by using the GOSim package [25]. The results, shown in figure 3, indicate that for all the three GO branches, recent duplicates tend to be more similar than older ones. Indeed, average GO term similarity values tend to decrease as the duplication time increases. On the other hand, the mean GO term similarity of duplicates in all duplication date group never reach values lower than one half, indicating that also ancient pre-WGD duplicates tend to maintain some functional overlap. The curve of GO similarity versus duplication age reaches lower values for the "biological process" and "cellular component" branches. This indicates that duplicates are more likely to diversify the biological process they participate into and the cellular compartment to which they belong rather than their molecular function. Secondly, they do so at domain score nearly fixed to a value close to one, indicating that on average the function of duplicates migrates within the same fold structure, presumably by sequence mutations or recombinations maintaining the same structural domains [26].

The same trends are also visible from the histograms of GOsim and domain-based similarity scores of all duplicate pairs (figure 4). The pairs of duplicates having high domain-based similarity is consistently higher in number than those with high GO similarity, but this trend is weaker for the "molecular function" taxonomy. In order to gather more direct evidence of this general domain and functional conservation under strong sequence evolution, we also compared these figures with the normalized histogram of the (protein) sequence identity ($\%id/100$) between pairs of duplicates from Smith-Waterman pairwise alignments, performed by using EMBOSS Water [27] (figure 4). The latter distribution has the lowest peak at one and the highest value at low scores, confirming that strong migration in protein sequence accompanies stability of domains and functions.

In order to exclude biases of computational nature that could influence the results, we repeated the analyses with different conditions. Firstly, not all proteins *S. cerevisiae* are covered entirely by domains, but some have gaps. Excluding from the analysis proteins with gaps should confirm that the functional migration of paralogs is not attributable to unknown domains. Supplementary figure S4 shows that this is indeed the case. Secondly, Gene Ontology annotations inferred from computational evidence could generate false positives in GO similarity, especially in the case of recent duplicates with significant sequence similarity. To circumvent this possibility, we restricted the analysis to manually curated genes. This filter reduces significantly our dataset, especially in the case of non-WGD duplicates. For this reason, we grouped non-WGD paralogs in two pre- and post-WGD sets. This gave sufficient statistics to retrieve the same trends of figure 3 for the "biological process" and "molecular function" GO branches (supplementary figure S3), but not in the case of the "cellular component" GO branch, where the data are insufficient.

## Functional connotation of WGD and non-WGD paralogs

Next, we focused on the difference in function between local and global duplicates. Whole-genome and local duplications are different biological processes, and the analysis of WGD and non-WGD paralogs can help understanding the biological constraints laying behind the different processes leading to long-term persistence of duplicated pairs in the two cases [24, 28, 29]. In particular, different works proved that WGD and non-WGD duplicates are enriched for different functional classes of genes. Thus, we set out to quantify with our methods how the effects of the WGD on the genome are qualitatively different from those brought by local duplications.

### Domain-based analysis

Functional assignment of domains can be used for evaluating the evolutionary destiny of duplicates. We considered two functional classifications for domains given in the SCOP database [6, 7]. We then proceeded to evaluate the trends in domain duplications, regardless from the specific protein they were duplicated with. We assigned domains to a set $\mathcal{O}$ if they were duplicated in at least one WGD paralog, and a set $\mathcal{P}$ if they appeared in at least one local duplication (see Methods). We considered paralogs the genes that are recognized by homology criterion **B** and do not belong to set $\mathcal{O}$. First, we found that the intersection of these two sets, in the universe of all SUPERFAMILY domains, is larger (P-value $< 10^{-28}$) than expected from a hypergeometric null model (figure 5). Thus, there is a dominant common set of domains that is prone to be duplicated, regardless of the local or global duplication mode.

On the other hand, the observed distribution of the fraction of WGD versus non-WGD duplicate proteins where each domain topology is found is very uneven (supplementary figure S5). This trend indicates the existence of two populations of domain topologies: those that are duplicated only outside the WGD, and those that appear in both kinds of duplications, but have a bias towards being found in the WGD only. Consequently, we analyzed the sets $\mathcal{O} \setminus \mathcal{P}$, the domains only found in WGD duplicates, and $\mathcal{P} \setminus \mathcal{O}$, the domains only found in non-WGD duplicates, for functional enrichment. For the finer categories of the SCOP functional classification we found a few cases where the enrichment was biased in two opposite ways in the two sets, i.e. categories having a positive $Z$-score for WGD domains, and a negative $Z$-score for non-WGD domains.

The categories that show a bias for WGD-specific domains (belonging to $\mathcal{O} \setminus \mathcal{P}$) correspond to functions that are growth-related (ribosomes, translation), involved in regulation of gene transcription and degradation (transcription factors, proteases), primary metabolism (coenzymes) or cell adhesion. On the other hand, a positive bias for locally duplicated domains (belonging to $\mathcal{P} \setminus \mathcal{O}$) was found in functional categories related to transport, post-transcriptional regulatory processes and secondary metabolism. Surprisingly, we found that the category DNA repair and replication tends to be enriched among domains duplicated locally rather than globally. Weaker signals for the same trend were found for RNA processing and modification, chromatin structure and dynamics, toxins and defense enzymes.

### Gene Ontology analysis

In parallel, we performed a more standard functional characterization based on Gene Ontology analysis on the proteins, along the lines of previous studies [28, 29]. We considered the disjoint sets of WGD and non-WGD paralogs. For each set we extracted the over-represented GO terms, and we compared them looking for the terms shared between WGD and non WGD-paralogs or specifically connected to a group (over-represented in a group and not significantly present in the other). WGD and non-WGD paralogs are enriched in different GO terms. We performed the same analysis also on randomized sets. Two randomly assorted sets tend to share more over-represented GO terms than WGD paralogs and non-WGD paralogs. These results are inverted considering the terms specific for each group: differently from the random assorted groups, WGD paralogs and non-WGD paralogs have many exclusive genes (see Supplementary Results), indicating that WGD and non-WGD paralogs carry out different functions.

In accordance with the domain-based analysis and with the previous hierarchical analysis derived from expression profiles and functional annotations [24, 28], we find that WGD paralogs are enriched for genes involved in "fundamental" processes such as for example, ribosomes and translation, regulation of cell cycle, regulation of developmental processes, sporulation, NADP metabolic process. On the other side the non-WGD paralogs are enriched for genes involved in "peripheral" processes such as transport, amino acid transmembrane transport, cellular wall, vitamin metabolism.

Finally, a recent study by Guan and coworkers [28] found that WGD duplicates are more likely to share interaction partners and biological functions than non-WGD duplicates. To confirm the latter result, we analyzed the distribution of the GO similarity normalized histograms for all the pairs of the two disjoint sets. Indeed, WGD paralogs result slightly more similar than non-WGD paralogs for all the three GO branches (supplementary figure S2). On the other hand, comparing with figure 3, one notices that pre-WGD paralogs are less similar at the functional level, so that this signal might come at least in part from the functional difference of ancient non-WGD paralogs.

# Discussion

Homology among distant paralogs and orthologs proteins is a difficult task because of sequence divergence. But it is well known that the structure of a protein is more conserved than its sequence. To score distant relationships among yeast and *K. waltii* proteins we used SCOP superfamilies domain assignments. This choice has three main reasons. First, these domains contain three-dimensional structural information, and are not solely based on sequence similarity, so that they can be considered, at least to a certain extent, "independent" from sequence alignments. Second, compared to the higher classification into "folds", they are defined to guarantee monophyly, excluding convergent evolution. Evolutionary information on domains is intrinsic of the classification scheme of the SCOP database, which is the basis for the hidden Markov models of the SUPERFAMILY database. Third, this choice was taken in previous studies [17, 30], which give a term of comparison.

The criteria and scores we used assume that two proteins derived from the same common ancestor if they have the same domain architecture, or a series of domains from the same protein families. This method allowed us to compare the more distant structural homology relationships with those obtained by sequence comparisons alone, and it also provided us with simple means to study the evolution of protein function from the structural viewpoint, at the genome-wide level. Naturally, the hidden Markov model assignment of domains depends on the scoring parameters. We limited our analysis to the criteria used by the SUPERFAMILY database [14, 17, 30]. A thorough analysis of the role of these parameters is presented in ref. [7].

**Domain architecture and homology.** Despite the sparse coverage of structural domains, it seems evident from our results that even elementary domain based homology criteria can recover most of the information obtained through sequence alignments techniques. Indeed, the criteria we defined are able to capture a large fraction of *Ensembl-Compara* homology classes, and behave similarly for local duplications or the WGD. On the other hand, the opposite is not true. Several domain-based homology relationships are not found by sequence alignment methods. We quantified this by measuring the fraction of domain-based homology classes not containing *Ensembl-Compara* classes. Criteria **A** and **B** have a similar percentage of homologs not detected in Ensembl, while criterion **C**, follows a different trend. This last criterion is the only one that allows for insertion of external domains after duplication, which is an event that has been observed and can be expected from our knowledge of the evolutionary dynamics of proteins [31, 32, 33].

On the other hand, the different behavior of criterion **C** could suggest a lower reliability compared to the other ones. It is important to stress that the architecture comparison methods implemented in this paper can show false-positive matches. In other words, the less restrictive the criterion is, the higher is the possibility to incorrectly identify evolutionarily unrelated genes as homologs.

Overall, while some instances could represent false positives, we believe it is natural to expect that some others represent distant relationships that are not detected as paralogs by sequence alignment methods, but are recognized by domain-based methods. Our tools do not allow to quantify these false positives directly. However, we have accessed some other observables that go in favor of the reliability of domain-based criteria. Firstly the mean domain scores and, to a certain extent, the mean architecture scores of duplicates are very close to one, and remain invariant with duplication age (figures 2). This indicates that even ancient paralogs tend to have very similar domain composition. The slight drop of the architecture score for ancient pre-WGD duplicates suggests that even if paralogs tend to maintain their domain composition, the domain order or the number of repetitions may vary. Secondly, WGD paralogs do not show any peculiarities at the domain level compared to local duplicates and single-copy orthologs. On the contrary, there exists a significantly large set of "duplicable" domains, shared by the two duplication modes. This is in contrast with the markedly decreasing trend followed by the GO similarity score between paralogs as a function of duplication age, indicating that domains remain stable as protein function and sequence drift. Together, these data show that both the domains composing a protein, and the domain architectures are rather stable and independent from the specific evolutionary history, which goes in favor of homology criteria based on this aspect.

Thus, the above evidence goes in favor of using structural domains as a simple and computationally effective tool to discover gene duplications. At the same time, it points to some limitations of these methods. The most important of these is that currently no tool is available to quantify the failure rate of domain-based methods in detecting gene duplications. In other words, it would be important to estimate precisely which fraction of paralogs detected by domain-based methods and not by sequence alignment are really significant. For example, one cannot exclude that genes gained by horizontal transfer give rise to proteins with the same domain structure as some other proteins in the genome [34], or that the partial coverage of domain databases does not enable to resolve distinct architectures. However, an exact quantification of these processes is lacking.

**Domain structure and function of duplicate proteins.** A second and more biological question is to use domain architectures to understand gene duplication, and in particular the differences between local duplications and the WGD. To approach this question, we compared the results of our domain similarity scores with a functional evaluation at the level of both domains and genes.

Following duplications, proteins show divergence in their domain architectures. Our scoring criteria quantify the rate of divergence of architectures. For all duplications, the already mentioned fact that domain scores remain constant and close to one as a function of duplication age indicates a strong trend of conserving the domain composition. This has to be compared with the GO similarity analysis *on the same sets of duplicates* showing a marked trend for divergence in function with increasing duplication age. An explanation of this phenomenon may be the fact that proteins evolve with point mutations affecting one nucleotide at a time. Domain topology can withstand these mutations without changing significantly, but some elementary biochemical properties that define protein function may vary. In other words, point mutation can change protein function without changing their domain composition. It is well known that proteins with identical folds can diverge greatly not only in sequence but also in function [26].

Obviously, this functional divergence cannot exceed the physical possibilities of a domain topology: a kinase domain will never bind to DNA. This is compatible with our observation that GO similarities do not drop to zero, and even very ancient duplicates always retain some degree of functional overlap. Along the same lines, Wapinski and collaborators [24] observe that the functional fates of duplicates rarely diverge with respect to biochemical function, but typically diverge with respect to regulatory control. The typical case when this is known to happen is that of transcription factors [35], where the migration of sequences within the same DNA-binding fold can lead to major changes in the affinity for a given set of sequences, and thus to large variation on the set of regulated targets. More simply, GO term divergence could come to a change of cellular compartment or biological process while performing similar biochemical functions. Also note that the trend of the *Molecular function* GO taxonomy paralog similarity score with duplication age is weaker than the other two taxonomies, *Biological Process* and *Cel-* *lular Component*. We extracted from our set some paralogs that maintain exactly the same domain architecture after duplication, while changing their molecular function, their cellular compartment and/or the biological process in which they are involved (GO term similarity $< 0.15$). It is the case of BDH1 and SOR1, ancient pre-WGD duplicates (datation I). The first is a butanediol dehydrogenase involved in alcohol metabolic processes, while the second is a sorbitol dehydrogenase involved in hexose metabolism. SOR1 is also a post WGD duplicate (datation E) of XYL2, which encodes for a xylitol dehydrogenase. DIN7 and EXO1 are WGD duplicates, both encoding proteins with nuclease activity involved in DNA repair and replication. However, the first one is mitochondrial and the second is nuclear. Similarly the WGD paralogs SEC14 and YKL091C are both phosphatidylinositol/phosphatidylcholine transfer proteins, but the first performs its function in the cytosol and in the Golgi apparatus while the second is nuclear.

Naturally, the coverage of domains on genomes is only partial, which leaves the question open of whether the observed trends of functional annotations with duplication age are due to modifications in the space of domains that are not visible to our methods. While of course this may happen, it seems unlikely that this can affect the global observed trends, assuming that we are observing an unbiased random sample of the existing structural domains. In other words, if the domains that change their topology during evolution have a fixed probability to be in the set of known domains, this would generate on average a decreasing trend of the domain score with duplication age, which we do not observe. A confirmation of this is given by the fact that removing proteins with gaps (protein sequences of 100 or more aminoacids without an attribution of domain), all the observed trends (figure 4, supplementary figure S4) do not change.

**Specificity of the Whole-Genome Duplication.** We now revert to the specific features of the whole-genome duplication. Double-sided domain architecture comparison of *S. cerevisiae* WGD paralogs with their *K. waltii* ortholog allows to evaluate asymmetric evolution at the domain level. Comparing with a suitable null model, we found no systematic trend for asymmetry (table 2). This is not unexpected, as domains are much more stable than sequences in evolution, so that, even in presence of accelerated evolution at the sequence level, the fold structure could be conserved.

From the functional viewpoint, we observe that the WGD does not follow a different trend in GO-term similarity between paralogs than expected from its age. Thus, we have to conclude that a "functional burst" correlated to accelerated evolution [20] does not differentiate the global duplication from local ones, or that this trend is not visible from the data available to us.

Partitioning the universe of all *S. cerevisiae* domains in locally and globally duplicated ones yields two sets of WGD and non-WGD domains, that can have an intersection, as the same domain can be present in both WGD and non-WGD duplicates. Notably. this intersection is enormously larger than expected from a hypergeometric null model, which can be interpreted as the fact that, within the universe of domains, the main distinction is between domains found or not found in duplications, rather than between domains found in global versus local duplications. Thus, again, whole-genome and local duplications are unified, rather than separated by this trend.

However, the domains of WGD duplicates laying outside common set of duplicable domains remain significant, as they give rise to evident peaks in the frequency of observing a domain in the sets of WGD and non-WGD duplicates. Moreover, they are also significant functionally. Indeed, the disjoint sets of WGD-specific and local-duplication specific domains are enriched for different functional categories. Similar categories are found with a more standard functional analysis on the genes. The domain-based and the Gene Ontology functional analyses agree in underlining functional differences between WGD and non-WGD paralogs. There are several works that proved that WGD paralogs and non-WGD paralogs are similarly biased with respect to codon bias and evolutionary rate, although differing significantly in their functional constituency and in the medium number of interacting partners [24, 28, 29]. In agreement with these results, we find that fundamental functions, such as ribosomes and translation are enriched in the WGD while peripheral functions, such as secondary metabolism are enriched for local duplications. The rationale for this result might be that functions related to core biological processes, or in general realized by genes with more entangled genetic interactions are more difficult to replicate by duplicating one part at a time as it happens with local duplications [24]. On the other hand, global moves such as the WGD could release these constraints and allow “recycling” and disentanglement of more elaborate cell machinery.

Finally, we can speculate on the consequences of the fact that the functional dichotomy is also found at the domain level. If it is true that function migrates abundantly, the functional dichotomy of local and global duplicates may emerge from migration of function maintaining similar domain structures. However, this cannot be the only source of differentiation, because in that case the same functional differences would not emerge also from the analysis of WGD and non-WGD specific domains. On the contrary, our result indicate that the dichotomy must be at least in part a result of the “special” protein domains that are only found in local or global duplications.

## Methods

**Data Sets.** We used the SUPERFAMILY database version 1.69 [7, 15] for the SCOP superfamily domains assignment, and the functional annotation of domains. We implemented a C code to reconstruct the protein domain architectures, as ordered lists of domains and "gaps" (a protein subsequence of 100 AA or more not scored for domains). As a reference for homology assignment we used different homology tools based on sequence alignment and synteny. For sequence-based homology, we referred to *Ensembl-Compara* (release 50) [16]. For *K. waltii-S. cerevisiae* WGD duplicates we referred to refs. [8, 36] and to ref. [24]; the latter study was also used for the datation of duplicates.

**Homology criteria.** Three different homology criteria were used to compare the domain architecture of proteins [17, 32]. Criterion **A** considers exactly matching architectures. The underlying biological hypothesis is that divergence after duplication does not change the domain architecture of the proteins, implying that divergence between homologs should happen at the sequence/peptide level. Criterion **B** relaxes the previous condition, and considers homologous domain architectures that are equal or contain multiple repetition of ordered sets of domains, ignoring possible gap mismatches. Criterion **C** further relaxes the above conditions , considering domain architectures as homologous if one contains repeated architecture domain sequences possibly interspaced by gaps *or* other domains. The code that implements the three criteria is available from the authors upon request.

**Domain architecture comparison scores.** We defined two different methods to compare proteins in their structural properties. The first "domain score" quantifies the variation in the domains of the two architectures, and is defined as the number of common domains domains between the two architectures, divided by the total number of distinct domains found in both. The domain score measures the number of distinct domain topologies common to the two compared genes, ignoring gaps. It normalizes the score over the total number of different domains contained in the two architectures. The second "architecture score" takes into account the order of appearance of domains in the two architectures and is defined as the length of the longest matching string of domains and gaps between the two architectures, divided by their mean length. The architecture score measures the length (number of ordered domains) of the longest overlapping sequence between the domain architecture of the compared genes, treating gaps as domains (normalized over the mean length of the two architectures). Both scores have a range from 0 (no similarity) to 1 (full similarity). The scores for pairs of WGD, and non-WGD paralogs of different age groups were averaged and histogrammed. To test for asymmetric domain evolution of WGD duplicates, we considered a null model that randomly exchanges the values in the hash table containing the two scores between each of the *S. cerevisiae* paralogs and their corresponding *K. waltii* ortholog. The null hypothesis negates the anti-correlation expected in paralog proteins following uneven evolution. The code that implements the two scores is available with the authors.

**Domain-based functional analysis.** Duplicate proteins with nonempty domain architecture were divided into two disjoint sets of WGD and non-WGD duplicates. The first set, from ref. [36], is composed by 692 *S. cerevisiae* proteins, estimated to be 62% of the total WGD paralogs. The second set (1863 proteins) was defined by those proteins coded by a gene with at least one known homolog, from which we removed the other set. Structural domains extracted from the two sets were divided accordingly into three sets: the set $\mathcal{O}$ of domains found in WGD duplicates; the set $\mathcal{P}$ of domains found in non-WGD duplicates; the set $\mathcal{O} \cap \mathcal{P}$ of domains found in at least one member of both protein sets (figure 5). To assess the functional enrichment for WGD and non-WGD paralogs, we implemented a null model based on the hypergeometric distribution, which provides the expected number of domains assigned with function $\mathcal{F}$ belonging either to WGD paralogs or to non-WGD paralogs, using as universe the set of all distinct domains found in *S. cerevisiae.*

**Gene Ontology analysis** We downloaded the Gene Ontology (GO) annotation DAGs from the GO website (`http://www.geneontology.org`) and the gene product annotations from the Ensembl database, version 46. We considered a gene annotated to a GO term if it was directly annotated to it or to any of its descendants in the GO tree. We used the SYNERGY algorithm [24] for defining paralogy classes. Orthologs and paralogs were considered different groups. As a reference, 100 pairs of sets were considered, each consisting of 1000 randomly assorted genes with the only constraint that each gene was chosen only once in each pair. For each group we implemented an exact Fisher's test to assess whether a set of genes could be enriched in a certain GO term [37, 38]. Fisher's test gives the probability $P$ of obtaining an equal or greater number of genes annotated to the term in a set made of the same number of genes, but randomly selected. Subsequently, the terms shared by both groups and the exclusive terms (terms present in only one group) were extracted. Finally, we filtered the results retaining only GO terms with P-values $<= 10^{-3}$. For each pair of paralogs, we calculated the Lin GO term similarity, by using the *GOSim* R-package (Version 1.1.5.1) [25]. For each duplication date group we calculated the mean and the standard deviation of the mean of the GO term similarity.

## Acknowledgement

We would like to thank Hervé Isambert for useful discussions, and Paolo Provero for critical reading of this manuscript.

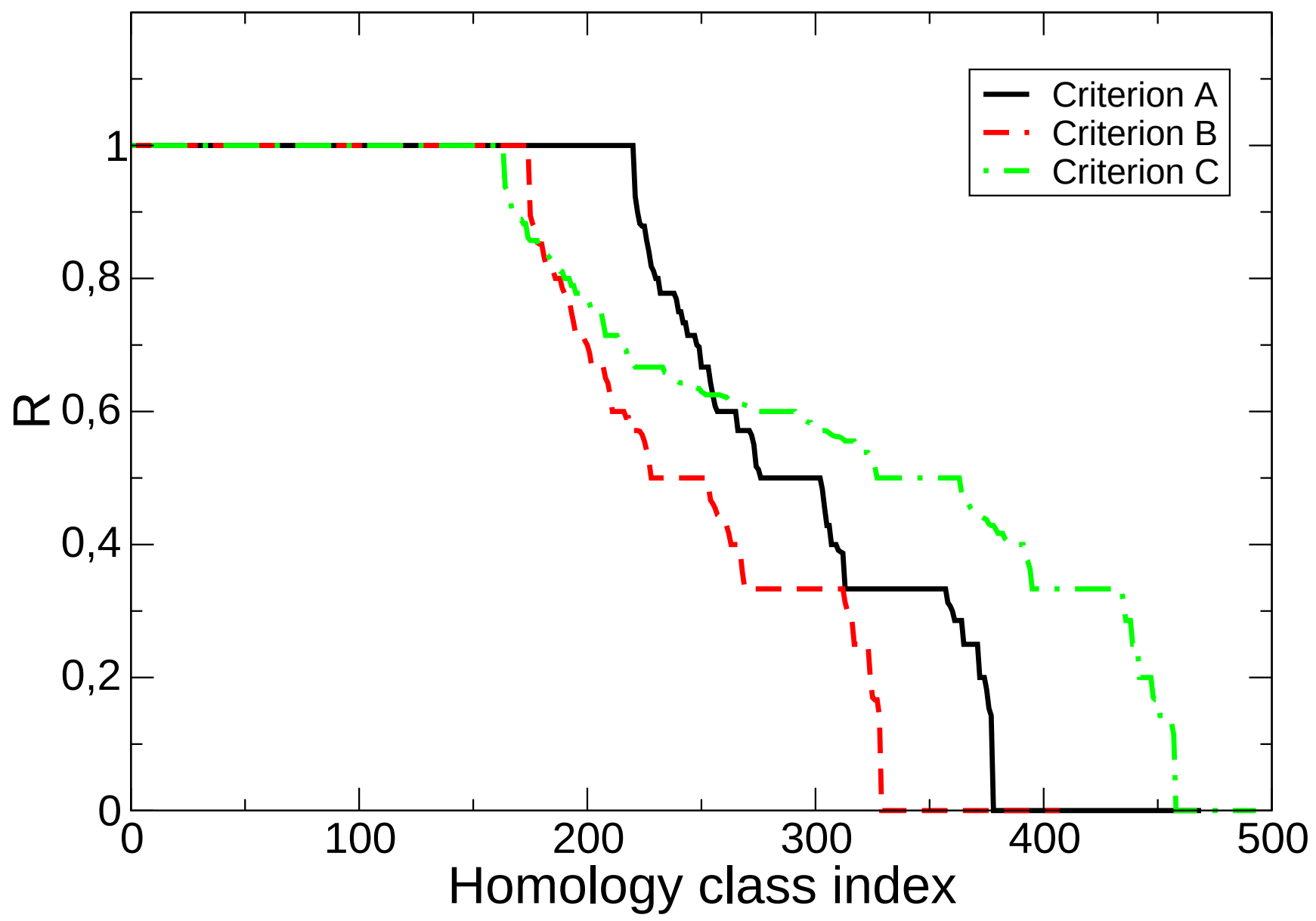

Figure 1: **The fraction of architecture homology classes R not recognized by *Ensembl-Compara***, plotted for all homology classes, ranked by R on the $x$-axis. The different lines in the plot refer to homology criteria **A** (black solid line), **B** (red dashed line) **C** (green dash-dotted line), defined in the text. Criterion **A** has the highest number of classes that are entirely not covered by *Ensembl-Compara* classes. Criterion **C**, while having the the lowest number of entirely not covered classes, also has the lowest rate of entirely covered ones.

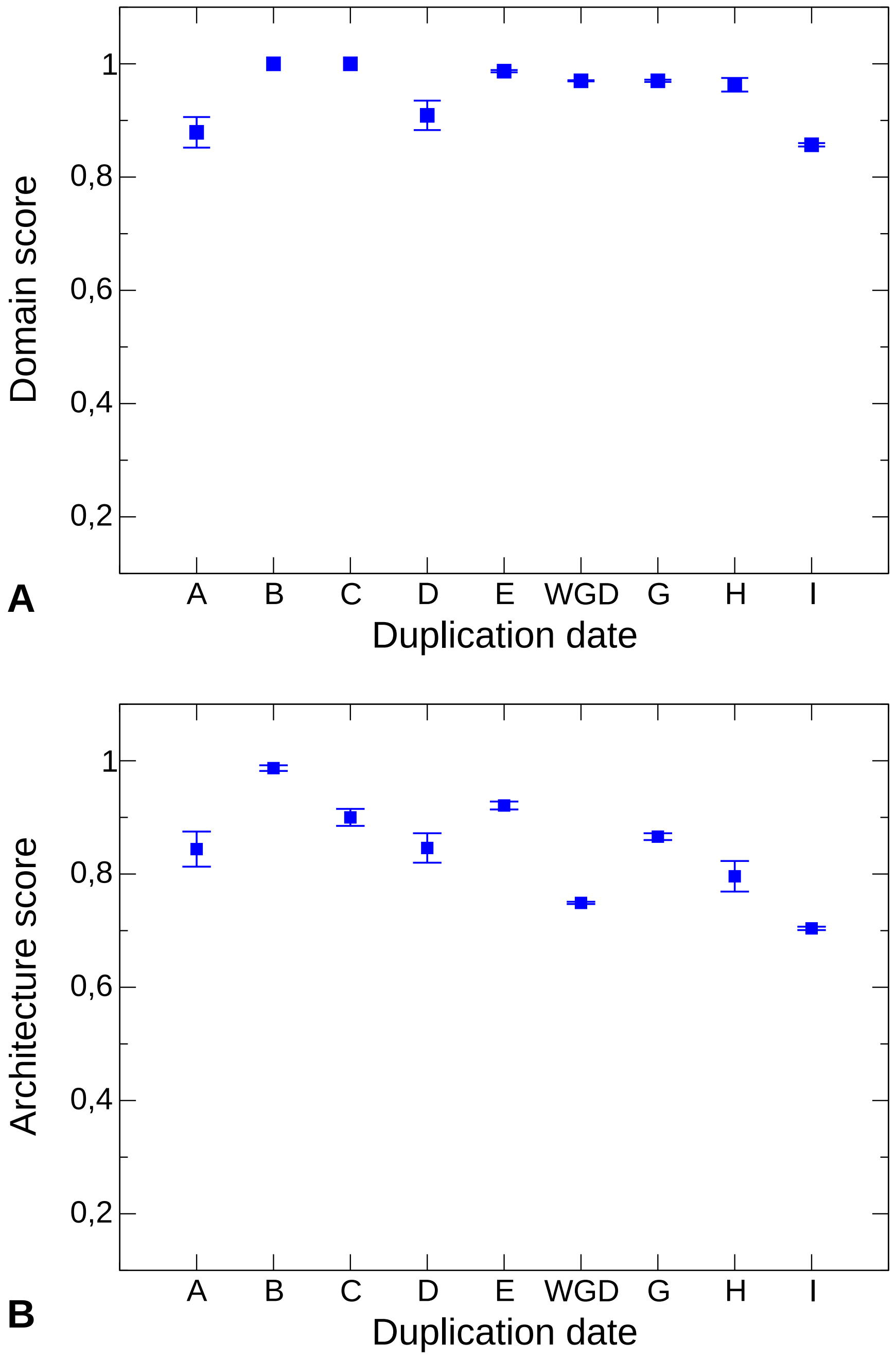

Figure 2: **A: Domain score as a function of duplication date. B: Architecture score as a function of duplication date.** The duplication refer to SYNERGY duplication age groups from [24]. A, B, C, D and E are post-WGD duplications, while I, H and G are pre-WGD duplications. The value for every date represent the mean over the scores of all duplicate pairs referring to that duplication age group.

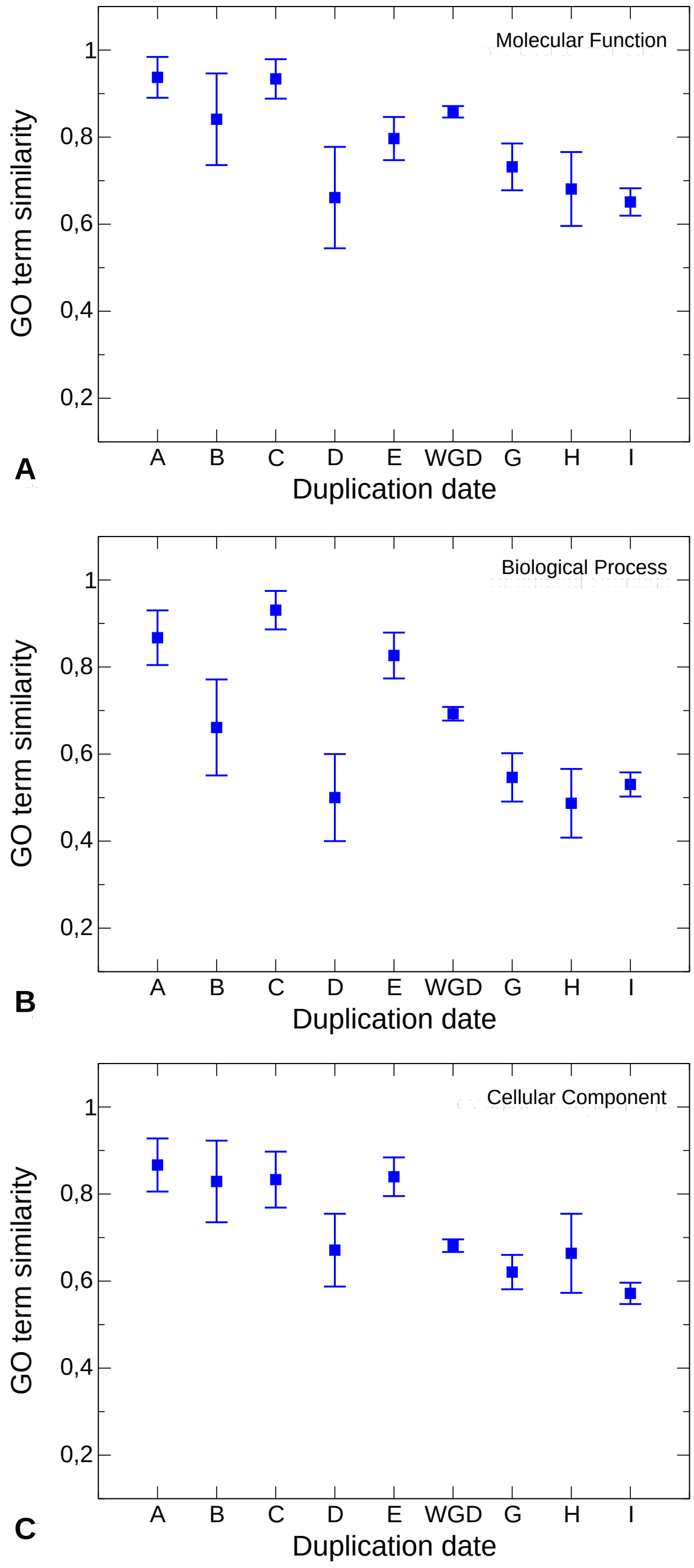

Figure 3: **Functional similarity of duplicates and duplication age.** The $y$-axes of the plots report mean similarity score (squares) and standard deviation (error bars) between the associated GO terms [25] of duplicates, computed over sets of duplicate pairs belonging to the same age groups ($x$-axis). The three panels refer to each of the three GO branches: molecular function (A), cellular component (B), biological process (C). Note that in all the plots the GO term similarity values tend to decrease with duplication age.

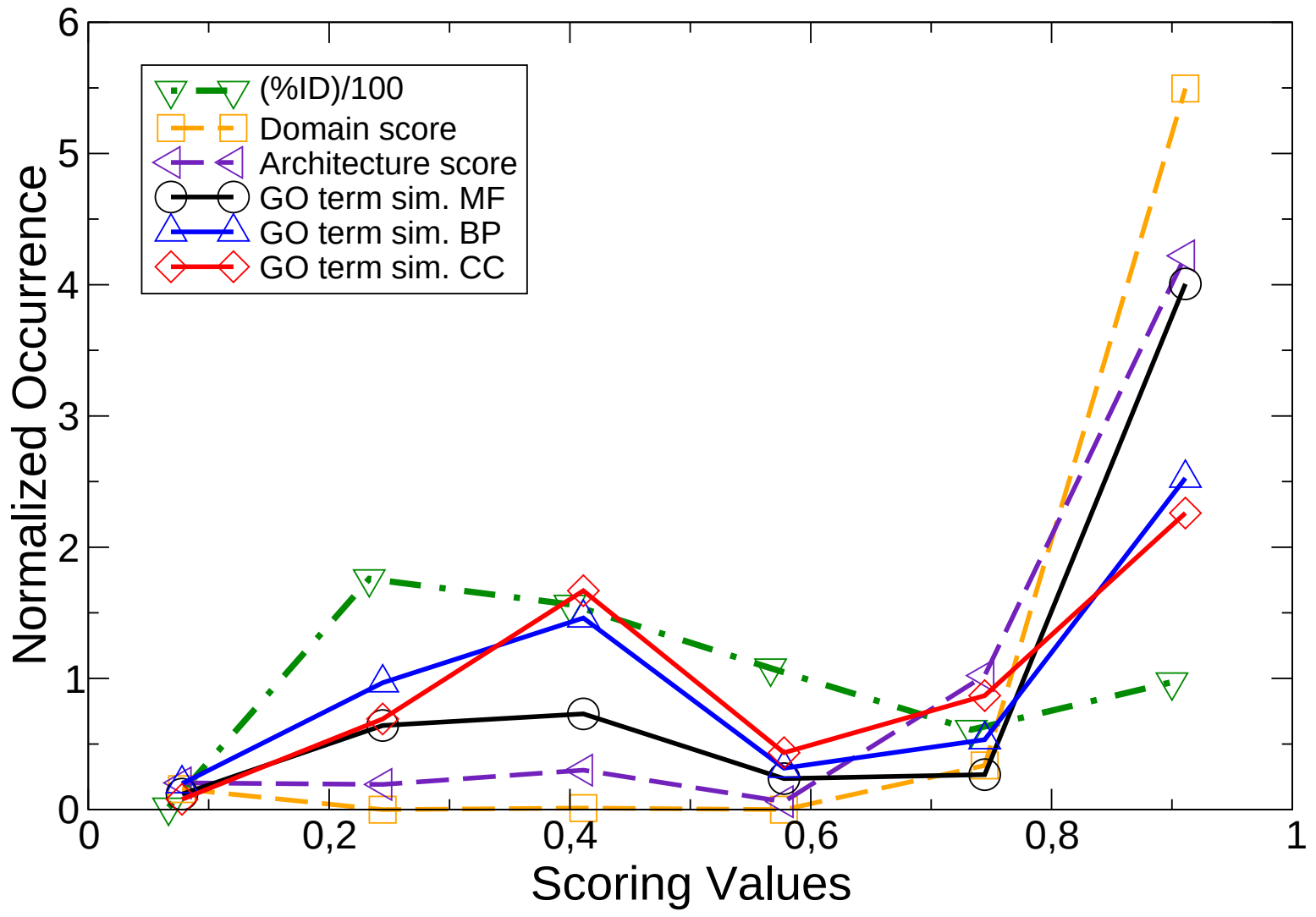

Figure 4: **Structural and Functional Divergence of Paralogs.** The plot reports histogram over all paralog pairs of domain score (squares), architecture score (left triangles), sequence identity (down triangles) and GO similarity (for all three taxonomies: molecular function, circles, biological process, up triangles, cellular compartment, diamonds). All curves are peaked around the value one, but the highest density values are reached by domain and architecture score curves, while the GO similarity reach lower values at one and develop a secondary peak below 0.5. This indicates that duplicates tend to maintain domain composition and architecture changing their functions.

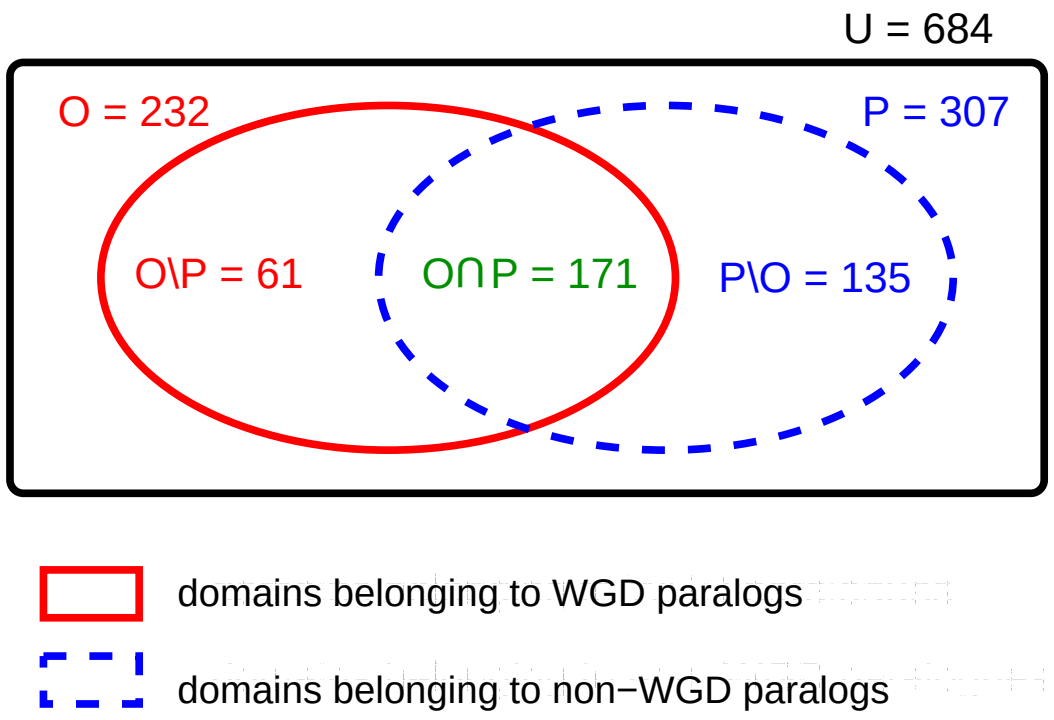

**A**

| Enriched in non−WGD (zP,zO) | Enriched in WGD (zP,zO) |
|---|---|
| Secondary metabolites transport (1.53 , −0,83) | Coenzyme metabolism (−1.92 , 1.09) |
| Other regulatory functions (1.02 , −0,94) | Ribosomes (−0.49 , 1.86) |
| Nitrogen metabolism (2.01 , −0,31) | Transcription factors (−1.24 , 0.34) |
| DNA replication and repair (1.31 , −1.78) | Cell adhesion (−1.32 , 0.50) |

**B**

Figure 5: **A: Venn diagrams of the sets involved in the domain-based functional enrichment analysis.** The empirical intersection is 11 standard deviations larger than the mean value provided by an hypergeometric distribution. **B: Table summarizing the significantly enriched functional classes for the sets of WGD and non-WGD domains.** zP and zO refer respectively to the Z-score for the non-WGD paralogs analysis and the WGD paralogs analysis (the sets $O \setminus P$ and $P \setminus O$ in panel A).

| Homology criterion | Triplets | (%) over total | Pairs | (%) over total |
|---|---|---|---|---|
| Kellis et al. | 457 | - | 2609 | - |
| Overlap | 289 | 100% | 1099 | 100% |
| Criterion A | 207 | 72% | 734 | 67% |
| Criterion B | 239 | 83% | 836 | 76% |
| Criterion C | 270 | 92% | 1010 | 91% |

| Homology criterion | *Ensembl* *Compara* classes | (%) over total |
|---|---|---|
| *Ensembl* | 672 | - |
| Overlap | 470 | 100% |
| Criterion A | 301 | 64% |
| Criterion B | 347 | 74% |
| Criterion C | 403 | 86% |

Table 1: **Comparison of classes obtained with domain-based homology criteria and homology classes built with WGD duplicates and their orthologs [8] (upper panel) and paralogs relations provided by *Ensembl-Compara* (lower panel,[16]).** For both tables, the first row of the shows the number of genes in sequenced-based homology classes. The second row reads the result of the intersection of these data with the architecture databases. The following three rows report the total and the relative fraction of the number of triplets and pairs found in the homology classes with criteria **A**, **B**, and **C**.

| Domain Score Comparison | | | | |
|---|---|---|---|---|
| | Empirical | Randomized | Difference (%) | P-value |
| F2 triplets | 80% | 73% | -7% | $< 10^{-4}$ |
| F1 triplets | 11% | 15% | -4% | $< 10^{-4}$ |
| **Architecture Score Comparison** | | | | |
| | Empirical | Randomized | Difference (%) | P-value |
| F2 triplets | 65% | 53% | -12% | $< 10^{-4}$ |
| F1 triplets | 16% | 50% | +34% | $< 10^{-4}$ |

Table 2: **Quantification of uneven architecture divergence between duplicates.** The table shows experimental and null-model relative frequencies of WGD paralogs in *S. cerevisiae* having identical architecture to their WGD ortholog in *K. waltii* according to the *domain* (upper panel) and *architecture* (lower panel) scores. The first two rows of each panel show the statistics restricted to the F2 and F1 triplets.

## Supporting Information

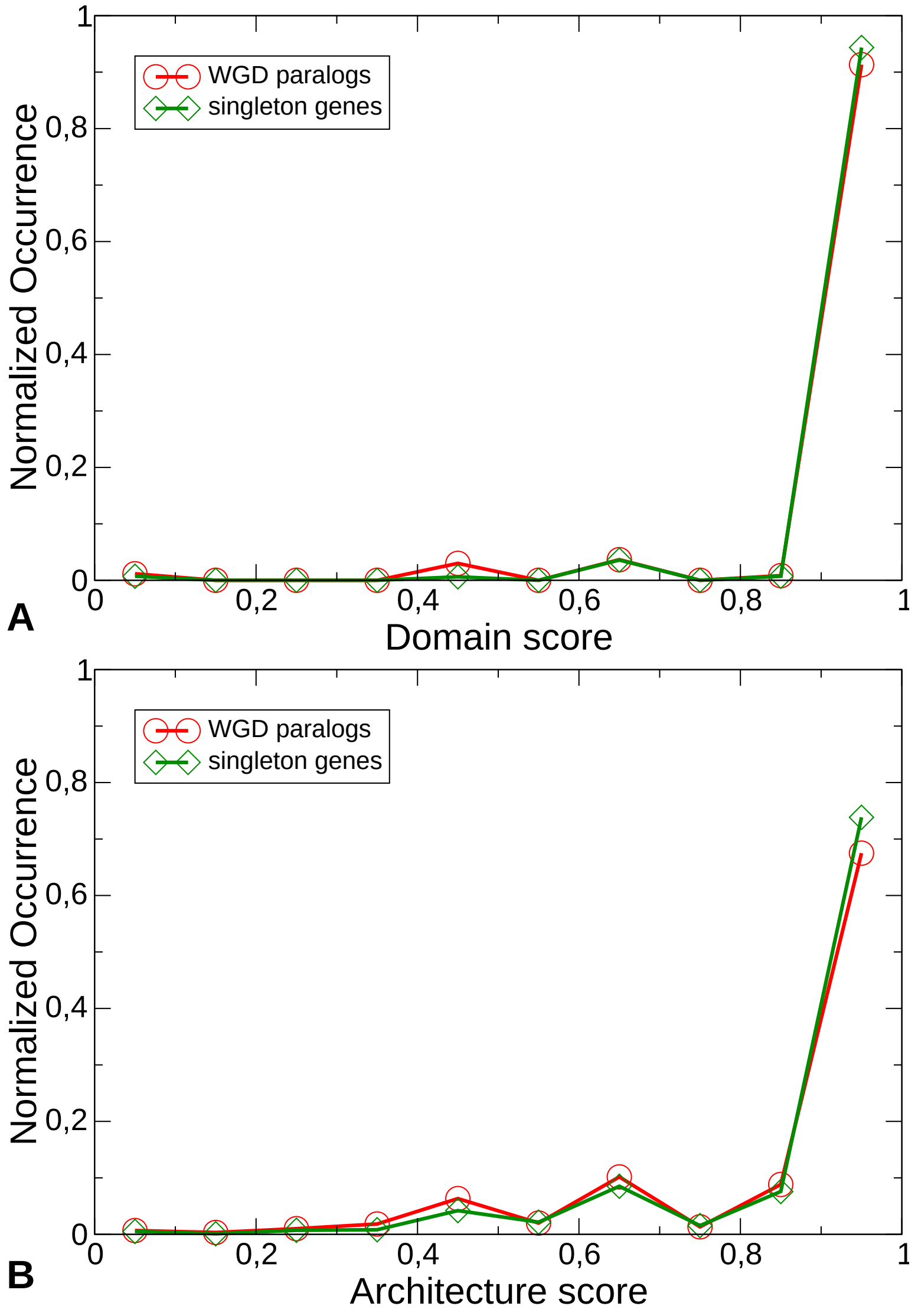

Figure S1: **WGD duplicates and single-copy orthologs show similar domain architecture divergence.** The plots report the histograms of domain score (A) and architecture score (B), evaluated in pairs of orthologs of *S. cerevisiae* and *K. waltii* for WGD duplicates and single copy S. cerevisiae-K. waltii orthologs.

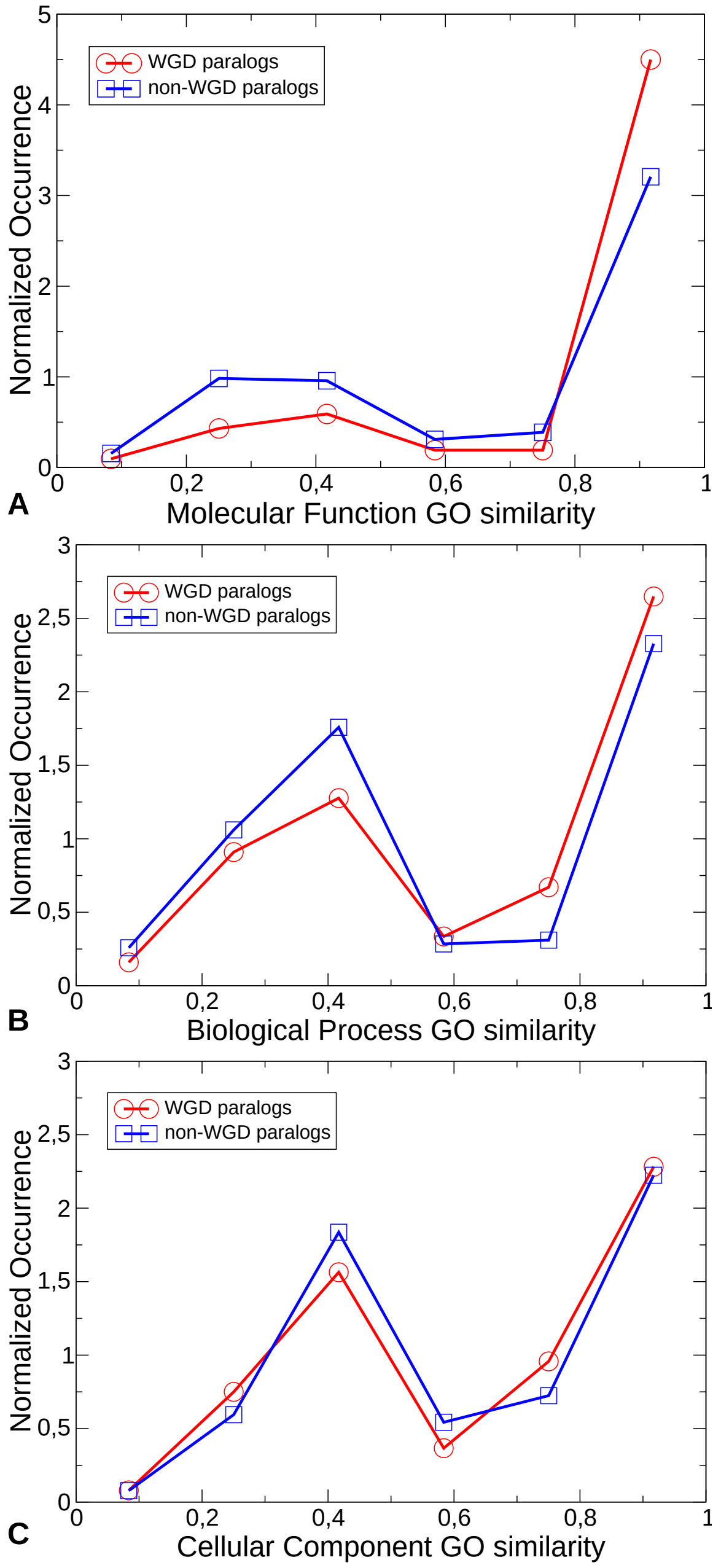

Figure S2: **Functional Similarity of WGD paralogs and non-WGD paralogs.** Normalized histograms of the Gene Ontology similarity between WGD and non-WGD duplicate pairs for the GO branches molecular function (A), biological process (B), cellular component (C). For all the three branches, WGD paralogs tend to have higher GO similarity scores than non-WGD paralogs.

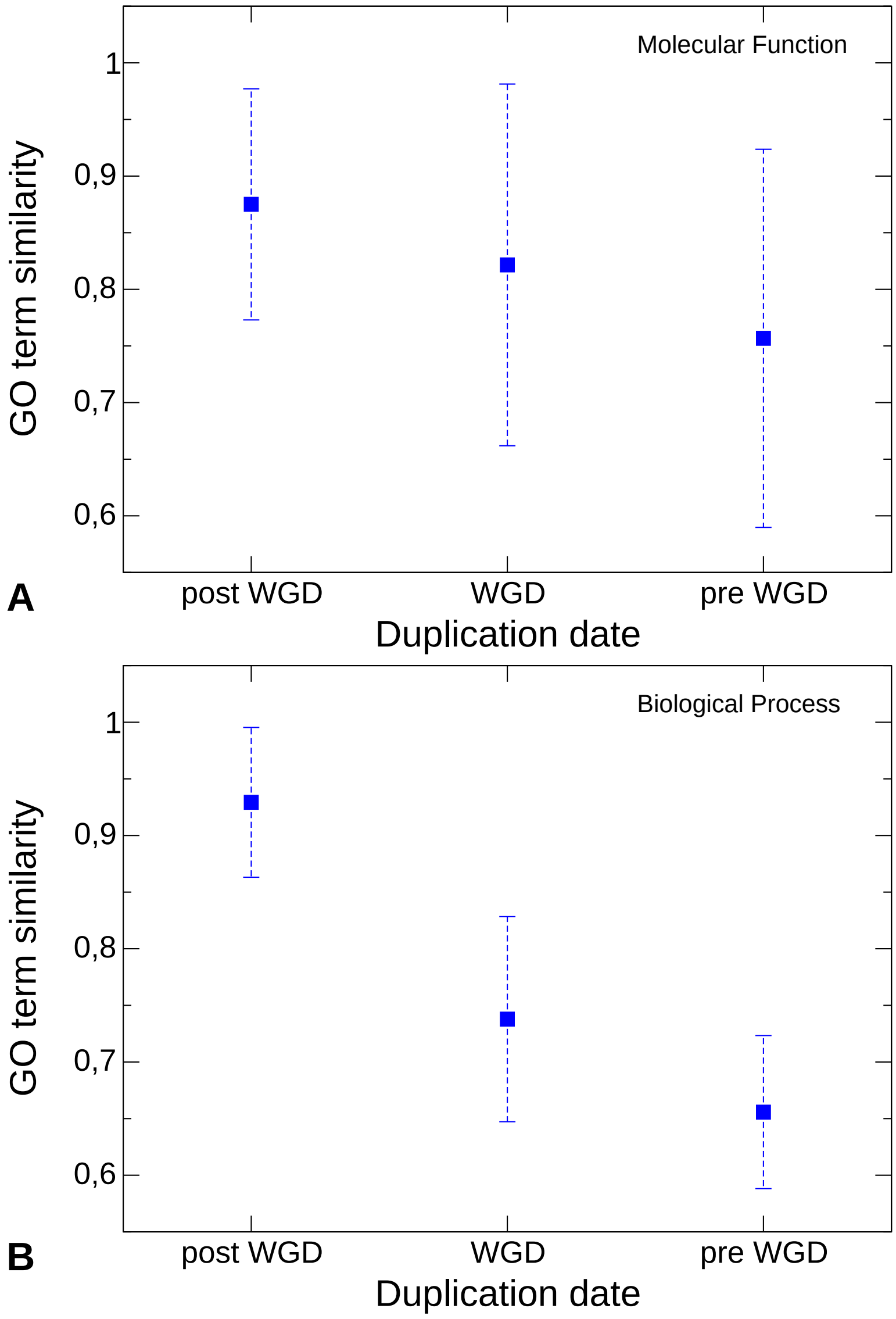

Figure S3: **Functional similarity of duplicates versus duplication age for manually curated GO annotations.** The plots report the mean (squares) and the standard deviation (error bars) of the GOsim similarity score between duplicates of the same age groups. The analysis was restricted only to the genes with experimental manually curated GO terms, grouping pre- and post-WGD duplication to gather sufficient statistics. This comparison is made for the GO branches: Biological Process (A), Molecular Function (B).

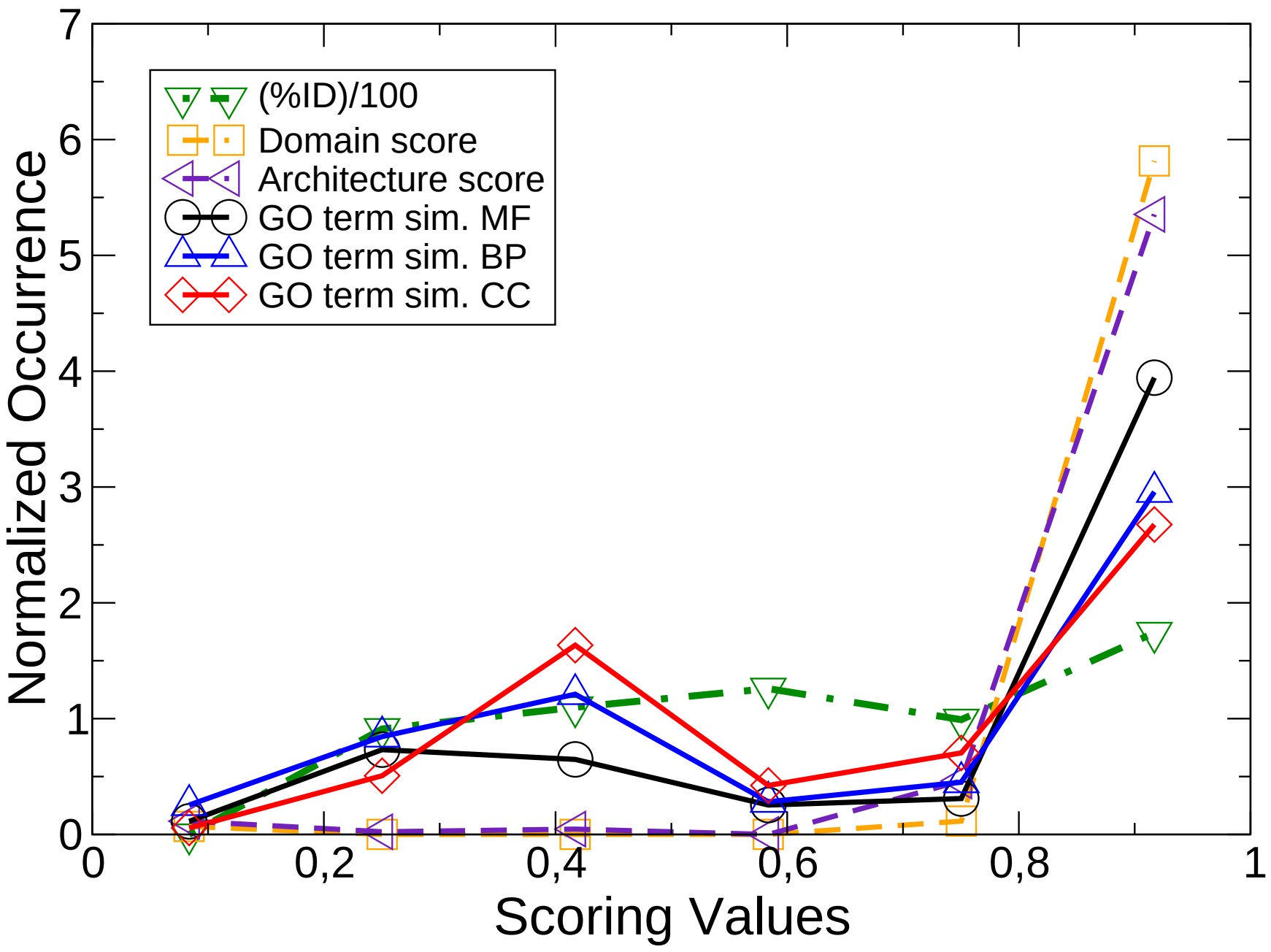

Figure S4: **Structural and functional divergence of paralogs with no gaps in the domain architecture.**. The plot reports histograms of sequence ID% retrieved from alignment, domain score, architecture score and GO term similarity (for all three branches) for all the paralog pairs with both proteins with by domain. Despite of this restriction we retrieve the same results shown in Figure 4 of the main text.

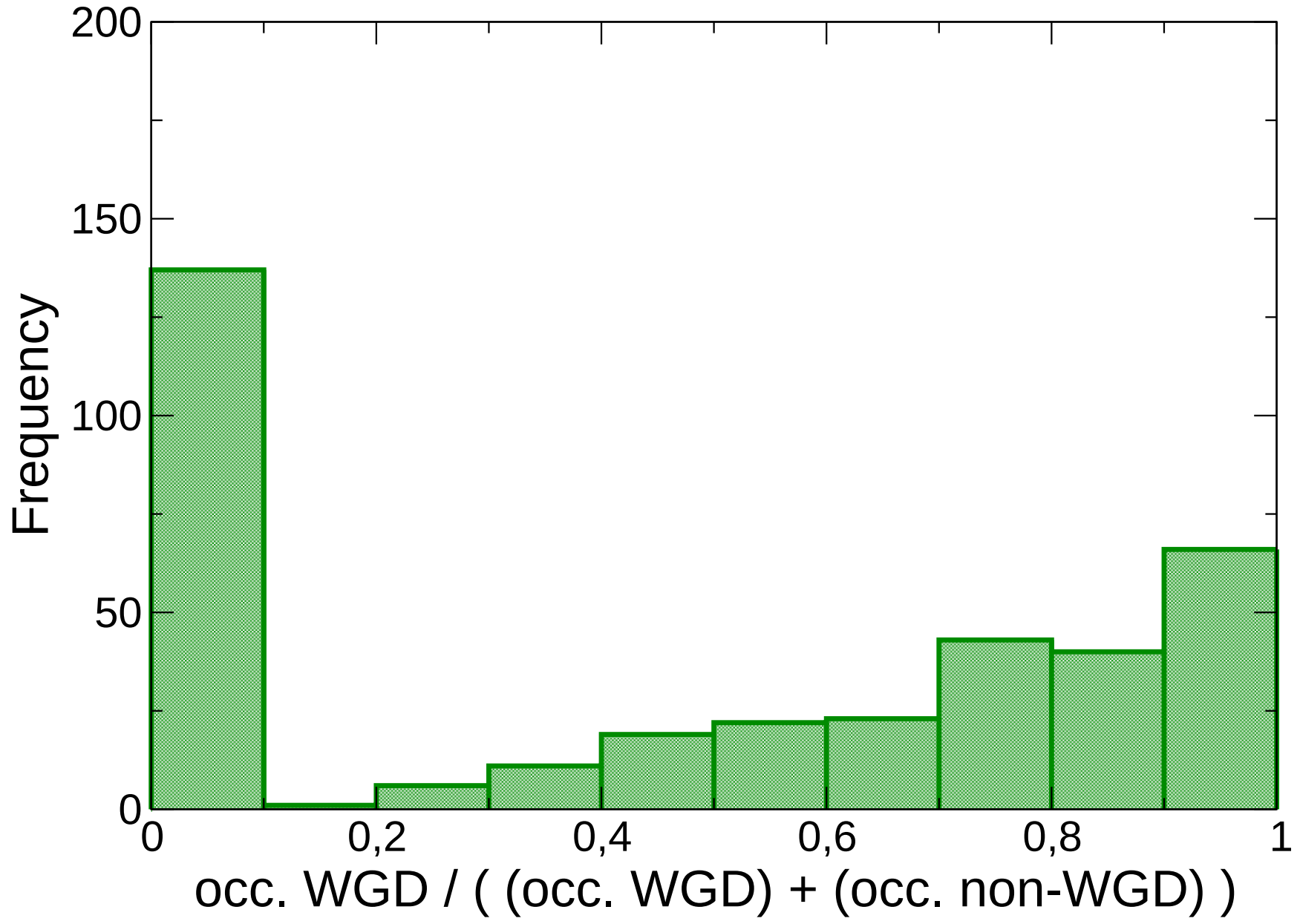

Figure S5: **Occurrence of domain topologies in WGD vs non-WGD duplicates.** For each SCOP domain, we calculated its occurrence in WGD proteins and non-WGD duplicates (normalized by the sizes of these two duplicate sets). The plot reports the histogram of the relative weight of occurrence of WGD duplicates, indicating the separation of two populations of domain topologies: domain topologies that appear in local duplications only (peak at zero), and those that appear in both the WGD and local duplications, having a preference towards the WGD (peak at one).

| Gene Ontology terms exclusive of WGD-Paralogs | | | |
|---|---|---|---|
| GO term | Number of genes | P-value | annotation |
| GO:0005737 | 571 | 3.62e-22 | cytoplasm |
| GO:0009987 | 647 | 1.72e-21 | cellular process |
| GO:0005622 | 675 | 8.80e-19 | intracellular |
| GO:0044424 | 668 | 1.10e-17 | intracellular part |
| GO:0005830 | 56 | 1.60e-17 | cytosolic ribosome (sensu Eukaryota) |
| GO:0005840 | 97 | 6.97e-16 | ribosome |
| GO:0005575 | 740 | 5.40e-15 | cellular component |
| GO:0005829 | 92 | 5.57e-15 | cytosol |
| GO:0044445 | 58 | 1.12e-14 | cytosolic part |
| GO:0044464 | 737 | 2.86e-14 | cell part |
| GO:0005623 | 737 | 3.11e-14 | cell |
| GO:0016773 | 62 | 4.53e-14 | phosphotransferase activity, alcohol group as acceptor |
| GO:0004674 | 49 | 5.75e-14 | protein serine/threonine kinase activity |
| GO:0009059 | 138 | 6.01e-14 | macromolecule biosynthetic process |
| GO:0004672 | 49 | 2.19e-13 | protein kinase activity |
| GO:0016301 | 66 | 4.33e-13 | kinase activity |
| GO:0003735 | 68 | 7.94e-13 | structural constituent of ribosome |
| GO:0009058 | 203 | 1.13e-12 | biosynthetic process |
| GO:0044262 | 69 | 1.45e-12 | cellular carbohydrate metabolic process |
| GO:0004713 | 42 | 3.62e-12 | protein-tyrosine kinase activity |
| GO:0065007 | 228 | 4.49e-12 | biological regulation |
| GO:0005488 | 536 | 6.77e-12 | binding |
| GO:0043284 | 31 | 7.92e-12 | biopolymer biosynthetic process |
| GO:0000271 | 25 | 7.93e-12 | polysaccharide biosynthetic process |
| GO:0006468 | 47 | 9.56e-12 | protein amino acid phosphorylation |
| GO:0044444 | 383 | 3.28e-11 | cytoplasmic part |
| GO:0007154 | 85 | 5.55e-11 | cell communication |
| GO:0007165 | 80 | 9.09e-11 | signal transduction |
| GO:0005843 | 26 | 1.60e-10 | cytosolic small ribosomal subunit (sensu Eukaryota) |
| GO:0006412 | 94 | 3.37e-10 | translation |
| GO:0032502 | 106 | 3.71e-10 | developmental process |
| GO:0016051 | 33 | 5.93e-10 | carbohydrate biosynthetic process |
| GO:0033279 | 56 | 1.01e-09 | ribosomal subunit |
| GO:0008152 | 520 | 1.6e-09 | metabolic process |
| GO:0050789 | 187 | 1.74e-09 | regulation of biological process |
| GO:0046164 | 27 | 2.35e-09 | alcohol catabolic process |
| GO:0006112 | 20 | 2.38e-09 | energy reserve metabolic process |
| GO:0044249 | 152 | 3.72e-09 | cellular biosynthetic process |
| GO:0044260 | 244 | 3.99e-09 | cellular macromolecule metabolic process |
| GO:0016052 | 30 | 5.11e-09 | carbohydrate catabolic process |
| GO:0044275 | 30 | 5.11e-09 | cellular carbohydrate catabolic process |
| GO:0050794 | 181 | 6.51e-09 | regulation of cellular process |
| GO:0016310 | 56 | 9.85e-09 | phosphorylation |
| GO:0005842 | 27 | 1.09e-08 | cytosolic large ribosomal subunit (sensu Eukaryota) |
| GO:0044237 | 485 | 1.35e-08 | cellular metabolic process |
| GO:0006739 | 13 | 1.38e-08 | NADP metabolic process |
| GO:0019320 | 24 | 1.41e-08 | hexose catabolic process |
| GO:0044264 | 27 | 1.55e-08 | cellular polysaccharide metabolic process |
| GO:0005976 | 27 | 1.55e-08 | polysaccharide metabolic process |
| GO:0044238 | 478 | 1.57e-08 | primary metabolic process |
| GO:0005516 | 11 | 1.86e-08 | calmodulin binding |
| GO:0032989 | 62 | 1.91e-08 | cellular structure morphogenesis |
| GO:0000902 | 62 | 1.91e-08 | cell morphogenesis |
| GO:0006007 | 23 | 2.22e-08 | glucose catabolic process |
| GO:0009250 | 16 | 2.50e-08 | glucan biosynthetic process |
| GO:0006006 | 30 | 2.56e-08 | glucose metabolic process |
| GO:0009653 | 62 | 2.63e-08 | anatomical structure morphogenesis |
| GO:0005198 | 81 | 2.95e-08 | structural molecule activity |
| GO:0005978 | 12 | 2.98e-08 | glycogen biosynthetic process |
| GO:0006796 | 65 | 3.81e-08 | phosphate metabolic process |
| GO:0006793 | 65 | 3.81e-08 | phosphorus metabolic process |
| GO:0006066 | 52 | 6.20e-08 | alcohol metabolic process |
| GO:0048856 | 62 | 7.67e-08 | anatomical structure development |
| GO:0007242 | 53 | 7.81e-08 | intracellular signaling cascade |
| GO:0046365 | 24 | 9.47e-08 | monosaccharide catabolic process |
| GO:0019318 | 34 | 9.49e-08 | hexose metabolic process |
| GO:0030529 | 107 | 1.25e-07 | ribonucleoprotein complex |
| GO:0006073 | 20 | 1.31e-07 | glucan metabolic process |
| GO:0007265 | 23 | 1.54e-07 | Ras protein signal transduction |
| GO:0005977 | 16 | 1.56e-07 | glycogen metabolic process |
| GO:0065008 | 74 | 1.59e-07 | regulation of biological quality |
| GO:0006740 | 11 | 1.78e-07 | NADPH regeneration |
| GO:0006897 | 28 | 2.25e-07 | endocytosis |
| GO:0010324 | 30 | 2.48e-07 | membrane invagination |
| GO:0019843 | 17 | 3.02e-07 | rRNA binding |
| GO:0050793 | 11 | 4.48e-07 | regulation of developmental process |
| GO:0016772 | 77 | 5.36e-07 | transferase activity, transferring phosphorus-containing groups |
| GO:0005933 | 40 | 6.06e-07 | cellular bud |
| GO:0005996 | 34 | 6.13e-07 | monosaccharide metabolic process |
| GO:0030955 | 9 | 7.98e-07 | potassium ion binding |
| GO:0051726 | 44 | 9.88e-07 | regulation of cell cycle |
| GO:0000074 | 44 | 9.88e-07 | regulation of progression through cell cycle |
| GO:0006098 | 10 | 1.03e-06 | pentose-phosphate shunt |
| GO:0009117 | 41 | 1.15e-06 | nucleotide metabolic process |
| GO:0007264 | 34 | 1.76e-06 | small GTPase mediated signal transduction |
| GO:0005979 | 6 | 2.99e-06 | regulation of glycogen biosynthetic process |
| GO:0051278 | 12 | 3.25e-06 | chitin- and beta-glucan-containing cell wall polysaccharide biosynthetic process |
| GO:0008360 | 8 | 5.04e-06 | regulation of cell shape |
| GO:0006038 | 8 | 5.04e-06 | cell wall chitin biosynthetic process |
| GO:0022603 | 8 | 5.04e-06 | regulation of anatomical structure morphogenesis |
| GO:0022604 | 8 | 5.04e-06 | regulation of cell morphogenesis |
| GO:0006769 | 17 | 5.74e-06 | nicotinamide metabolic process |
| GO:0044267 | 220 | 7.05e-06 | cellular protein metabolic process |
| GO:0015935 | 26 | 7.80e-06 | small ribosomal subunit |
| GO:0005935 | 31 | 8.82e-06 | cellular bud neck |
| GO:0019362 | 17 | 1.16e-05 | pyridine nucleotide metabolic process |
| GO:0006031 | 9 | 1.29e-05 | chitin biosynthetic process |
| GO:0006037 | 8 | 1.35e-05 | cell wall chitin metabolic process |
| GO:0000028 | 8 | 1.35e-05 | ribosomal small subunit assembly and maintenance |
| GO:0048610 | 36 | 1.53e-05 | reproductive cellular process |
| GO:0022413 | 36 | 1.53e-05 | reproductive process in single-celled organism |
| GO:0030427 | 37 | 1.59e-05 | site of polarized growth |
| GO:0016192 | 70 | 1.61e-05 | vesicle-mediated transport |
| GO:0005934 | 18 | 1.83e-05 | cellular bud tip |
| GO:0005498 | 6 | 1.88e-05 | sterol carrier activity |
| GO:0005496 | 6 | 1.88e-05 | steroid binding |
| GO:0032934 | 6 | 1.88e-05 | sterol binding |
| GO:0006887 | 17 | 2.22e-05 | exocytosis |
| GO:0015934 | 30 | 2.95e-05 | large ribosomal subunit |
| GO:0008361 | 33 | 3.01e-05 | regulation of cell size |
| GO:0015980 | 36 | 3.91e-05 | energy derivation by oxidation of organic compounds |
| GO:0009272 | 13 | 3.91e-05 | chitin- and beta-glucan-containing cell wall biogenesis |
| GO:0040007 | 34 | 4.31e-05 | growth |
| GO:0065009 | 21 | 4.50e-05 | regulation of a molecular function |
| GO:0042546 | 13 | 5.74e-05 | cell wall biogenesis |
| GO:0006665 | 12 | 6.26e-05 | sphingolipid metabolic process |
| GO:0010383 | 8 | 6.56e-05 | cell wall polysaccharide metabolic process |
| GO:0030011 | 6 | 6.75e-05 | maintenance of cell polarity |
| GO:0006869 | 14 | 7.15e-05 | lipid transport |
| GO:0050790 | 20 | 7.36e-05 | regulation of catalytic activity |
| GO:0031505 | 15 | 8.24e-05 | chitin- and beta-glucan-containing cell wall organization and biogenesis |
| GO:0006042 | 9 | 8.97e-05 | glucosamine biosynthetic process |
| GO:0006045 | 9 | 8.97e-05 | N-acetylglucosamine biosynthetic process |
| GO:0046349 | 9 | 8.97e-05 | amino sugar biosynthetic process |
| GO:0006893 | 12 | 9.31e-05 | Golgi to plasma membrane transport |

Table S1: **Gene Ontology terms exclusive of WGD paralogs.** The table reports the results of the enrichment analysis for Gene Ontology terms exclusive of non-WGD duplicates, with populations of functional categories (column two) and P-values from hypergeometric testing (column three).

| Gene Ontology terms exclusive of non-WGD paralogs | | | |
|---|---|---|---|
| GO term | Number of genes | P-value | annotation |
| GO:0022891 | 60 | 4.99e-16 | substrate-specific transmembrane transporter activity |
| GO:0022857 | 64 | 6.24e-16 | transmembrane transporter activity |
| GO:0022892 | 65 | 1.36e-13 | substrate-specific transporter activity |
| GO:0005215 | 71 | 2.38e-13 | transporter activity |
| GO:0005353 | 11 | 4.78e-11 | fructose transmembrane transporter activity |
| GO:0015578 | 11 | 4.78e-11 | mannose transmembrane transporter activity |
| GO:0005355 | 11 | 1.44e-10 | glucose transmembrane transporter activity |
| GO:0015149 | 11 | 3.86e-10 | hexose transmembrane transporter activity |
| GO:0015145 | 11 | 3.86e-10 | monosaccharide transmembrane transporter activity |
| GO:0015291 | 25 | 1.17e-09 | secondary active transmembrane transporter activity |
| GO:0015293 | 19 | 1.36e-09 | symporter activity |
| GO:0022804 | 35 | 3.71e-09 | active transmembrane transporter activity |
| GO:0015171 | 14 | 1.02e-08 | amino acid transmembrane transporter activity |
| GO:0015837 | 17 | 1.13e-08 | amine transport |
| GO:0051119 | 14 | 1.55e-08 | sugar transmembrane transporter activity |
| GO:0005351 | 14 | 1.55e-08 | sugar:hydrogen ion symporter activity |
| GO:0005342 | 19 | 1.83e-08 | organic acid transmembrane transporter activity |
| GO:0046943 | 18 | 3.04e-08 | carboxylic acid transmembrane transporter activity |
| GO:0015144 | 14 | 3.42e-08 | carbohydrate transmembrane transporter activity |
| GO:0006865 | 15 | 4.90e-08 | amino acid transport |
| GO:0046942 | 19 | 5e-08 | carboxylic acid transport |
| GO:0015849 | 19 | 6.35e-08 | organic acid transport |
| GO:0000023 | 8 | 7.87e-08 | maltose metabolic process |
| GO:0008615 | 8 | 7.87e-08 | pyridoxine biosynthetic process |
| GO:0042819 | 8 | 7.87e-08 | vitamin B6 biosynthetic process |
| GO:0008614 | 8 | 1.93e-07 | pyridoxine metabolic process |
| GO:0042816 | 8 | 1.94e-07 | vitamin B6 metabolic process |
| GO:0009277 | 19 | 1.42e-06 | chitin- and beta-glucan-containing cell wall |
| GO:0048503 | 13 | 3.21e-06 | GPI anchor binding |
| GO:0015205 | 6 | 9.08e-06 | nucleobase transmembrane transporter activity |
| GO:0015174 | 6 | 9.08e-06 | basic amino acid transmembrane transporter activity |
| GO:0042402 | 6 | 9.084e-06 | biogenic amine catabolic process |
| GO:0016020 | 168 | 1.22e-05 | membrane |
| GO:0005984 | 8 | 1.29e-05 | disaccharide metabolic process |
| GO:0015075 | 29 | 1.82e-05 | ion transmembrane transporter activity |
| GO:0042219 | 6 | 3.59e-05 | amino acid derivative catabolic process |
| GO:0015175 | 5 | 4.20e-05 | neutral amino acid transmembrane transporter activity |
| GO:0030976 | 5 | 4.20e-05 | thiamin pyrophosphate binding |
| GO:0019660 | 5 | 4.20e-05 | glycolytic fermentation |
| GO:0006559 | 4 | 6.82e-05 | L-phenylalanine catabolic process |
| GO:0031224 | 124 | 7.03e-05 | intrinsic to membrane |
| GO:0030287 | 5 | 8.98e-05 | cell wall-bounded periplasmic space |
| GO:0009083 | 5 | 8.98e-05 | branched chain family amino acid catabolic process |
| GO:0044270 | 9 | 9.37e-05 | nitrogen compound catabolic process |
| GO:0009310 | 9 | 9.37e-05 | amine catabolic process |
| GO:0016021 | 123 | 9.81e-05 | integral to membrane |

Table S2: **Gene Ontology terms exclusively found in non-WGD Paralogs.** The table reports the results of the enrichment analysis for Gene Ontology terms exclusive of non-WGD duplicates, with populations of functional categories (column two) and P-values from hypergeometric testing (column three).

| SCOP superfamily domain occurrence | | |
|---|---|---|
| Domain | Occurrence in WGD proteins | Occurrence in non-WGD proteins |
| 46561 | 2 | 0 |
| 46565 | 0 | 16 |
| 46579 | 0 | 7 |
| 46589 | 0 | 2 |
| 46626 | 2 | 0 |
| 46689 | 8 | 14 |
| 46774 | 0 | 2 |
| 46785 | 8 | 13 |
| 46906 | 2 | 0 |
| 46934 | 2 | 3 |
| 46938 | 2 | 2 |
| 46946 | 2 | 1 |
| 46955 | 0 | 2 |
| 46977 | 2 | 0 |
| 47060 | 0 | 2 |
| 47072 | 0 | 2 |
| 47095 | 4 | 3 |
| 47113 | 2 | 22 |
| 47212 | 2 | 0 |
| 47240 | 2 | 1 |
| 47323 | 2 | 2 |
| 47370 | 4 | 2 |
| 47459 | 0 | 8 |
| 47473 | 2 | 10 |
| 47576 | 0 | 2 |
| 47592 | 4 | 0 |
| 47616 | 0 | 5 |
| 47661 | 2 | 3 |
| 47672 | 1 | 0 |
| 47694 | 0 | 2 |
| 47769 | 2 | 2 |
| 47807 | 2 | 1 |
| 47819 | 0 | 2 |
| 47923 | 4 | 5 |
| 47954 | 10 | 10 |
| 47973 | 0 | 2 |
| 48019 | 0 | 4 |
| 48065 | 2 | 2 |
| 48097 | 0 | 2 |
| 48140 | 2 | 0 |
| 48150 | 2 | 1 |
| 48168 | 2 | 0 |
| 48179 | 2 | 5 |
| 48208 | 2 | 6 |
| 48225 | 0 | 2 |
| 48239 | 0 | 4 |
| 48256 | 2 | 1 |
| 48264 | 0 | 3 |
| 48317 | 2 | 4 |
| 48334 | 0 | 2 |
| 48350 | 6 | 4 |
| 48366 | 2 | 1 |
| 48371 | 8 | 57 |
| 48403 | 6 | 6 |
| 48425 | 2 | 2 |
| 48431 | 1 | 0 |
| 48439 | 0 | 6 |
| 48445 | 2 | 0 |
| 48452 | 6 | 24 |
| 48464 | 6 | 6 |
| 48557 | 0 | 3 |
| 48576 | 0 | 3 |
| 48592 | 0 | 6 |
| 48613 | 0 | 5 |
| 48695 | 2 | 0 |
| 49348 | 2 | 0 |
| 49354 | 2 | 0 |
| 49447 | 0 | 2 |
| 49493 | 0 | 2 |
| 49562 | 2 | 2 |
| 49764 | 0 | 3 |
| 49777 | 0 | 3 |
| 49785 | 0 | 3 |
| 49863 | 1 | 0 |
| 49879 | 6 | 4 |
| 49899 | 4 | 4 |
| 50044 | 9 | 11 |
| 50104 | 6 | 1 |
| 50129 | 0 | 4 |
| 50182 | 0 | 16 |
| 50193 | 2 | 1 |
| 50249 | 10 | 16 |
| 50324 | 0 | 2 |
| 50447 | 5 | 4 |
| 50465 | 3 | 2 |
| 50475 | 2 | 2 |
| 50630 | 2 | 10 |
| 50677 | 0 | 2 |
| 50729 | 12 | 8 |
| 50800 | 2 | 0 |
| 50891 | 4 | 3 |
| 50965 | 4 | 1 |
| 50978 | 9 | 83 |
| 50985 | 0 | 3 |
| 51011 | 0 | 7 |
| 51161 | 0 | 2 |
| 51182 | 0 | 3 |
| 51206 | 0 | 2 |
| 51230 | 4 | 2 |
| 51246 | 4 | 0 |
| 51306 | 0 | 3 |
| 51316 | 0 | 3 |
| 51366 | 2 | 5 |
| 51395 | 0 | 7 |
| 51412 | 2 | 4 |
| 51419 | 0 | 2 |
| 51430 | 2 | 14 |
| 51445 | 4 | 18 |
| 51556 | 1 | 5 |
| 51569 | 6 | 4 |
| 51604 | 2 | 3 |
| 51621 | 2 | 3 |
| 51645 | 0 | 2 |
| 51726 | 0 | 2 |
| 51730 | 0 | 3 |
| 51735 | 12 | 61 |
| 51905 | 10 | 10 |
| 51998 | 2 | 0 |
| 52016 | 0 | 4 |
| 52025 | 2 | 1 |
| 52047 | 2 | 6 |
| 52058 | 4 | 3 |
| 52080 | 2 | 2 |
| 52087 | 2 | 2 |
| 52096 | 4 | 0 |
| 52113 | 0 | 3 |
| 52151 | 4 | 3 |
| 52161 | 2 | 1 |
| 52166 | 2 | 1 |
| 52172 | 1 | 0 |
| 52218 | 2 | 2 |
| 52283 | 2 | 3 |
| 52313 | 2 | 1 |
| 52317 | 4 | 8 |
| 52335 | 2 | 0 |
| 52343 | 4 | 3 |
| 52374 | 4 | 11 |
| 52402 | 1 | 6 |
| 52440 | 4 | 0 |
| 52467 | 2 | 7 |
| 52490 | 2 | 2 |
| 52507 | 2 | 2 |
| 52518 | 2 | 6 |
| 52540 | 32 | 121 |
| 52743 | 2 | 0 |
| 52768 | 0 | 6 |
| 52777 | 0 | 4 |
| 52799 | 2 | 10 |
| 52821 | 2 | 6 |
| 52833 | 16 | 24 |
| 52922 | 2 | 0 |
| 52935 | 2 | 0 |
| 52949 | 0 | 2 |
| 52954 | 2 | 0 |
| 52972 | 0 | 2 |
| 53032 | 0 | 2 |
| 53067 | 12 | 23 |
| 53092 | 0 | 2 |
| 53098 | 2 | 54 |
| 53137 | 2 | 2 |
| 53167 | 0 | 2 |
| 53187 | 2 | 7 |
| 53223 | 0 | 4 |
| 53244 | 2 | 1 |
| 53254 | 4 | 13 |
| 53271 | 6 | 6 |
| 53328 | 0 | 2 |
| 53335 | 0 | 45 |
| 53383 | 4 | 30 |
| 53448 | 12 | 12 |
| 53474 | 9 | 31 |
| 53613 | 0 | 9 |
| 53623 | 2 | 1 |
| 53633 | 2 | 0 |
| 53649 | 2 | 4 |
| 53659 | 2 | 5 |
| 53686 | 0 | 6 |
| 53697 | 1 | 2 |
| 53720 | 0 | 11 |
| 53732 | 0 | 4 |
| 53738 | 2 | 1 |
| 53756 | 4 | 7 |
| 53774 | 2 | 5 |
| 53850 | 0 | 4 |
| 53901 | 0 | 4 |
| 53927 | 1 | 5 |
| 54001 | 6 | 17 |
| 54189 | 2 | 2 |
| 54197 | 3 | 4 |
| 54211 | 6 | 15 |
| 54236 | 4 | 12 |
| 54427 | 0 | 3 |
| 54495 | 2 | 13 |
| 54534 | 2 | 2 |
| 54570 | 0 | 2 |
| 54575 | 2 | 0 |
| 54616 | 2 | 0 |
| 54626 | 0 | 2 |
| 54631 | 2 | 2 |
| 54637 | 0 | 5 |
| 54686 | 0 | 2 |
| 54695 | 2 | 2 |
| 54747 | 2 | 0 |
| 54768 | 0 | 5 |
| 54791 | 0 | 3 |
| 54826 | 2 | 3 |
| 54843 | 2 | 1 |
| 54849 | 0 | 6 |
| 54897 | 2 | 2 |
| 54928 | 10 | 40 |
| 54980 | 2 | 0 |
| 54999 | 0 | 2 |
| 55021 | 2 | 2 |
| 55035 | 2 | 0 |
| 55060 | 2 | 3 |
| 55103 | 0 | 2 |
| 55120 | 4 | 4 |
| 55129 | 2 | 2 |
| 55154 | 2 | 0 |
| 55174 | 2 | 3 |
| 55190 | 2 | 0 |
| 55205 | 0 | 2 |
| 55257 | 0 | 4 |
| 55277 | 2 | 0 |
| 55282 | 2 | 1 |
| 55298 | 2 | 0 |
| 55307 | 2 | 2 |
| 55315 | 4 | 4 |
| 55424 | 2 | 1 |
| 55455 | 2 | 2 |
| 55469 | 0 | 2 |
| 55486 | 2 | 4 |
| 55608 | 0 | 6 |
| 55666 | 0 | 2 |
| 55681 | 2 | 5 |
| 55729 | 0 | 18 |
| 55753 | 0 | 3 |
| 55797 | 2 | 1 |
| 55811 | 0 | 7 |
| 55821 | 0 | 2 |
| 55856 | 0 | 5 |
| 55874 | 2 | 0 |
| 55920 | 0 | 6 |
| 55957 | 2 | 1 |
| 55973 | 2 | 0 |
| 55979 | 0 | 2 |
| 56019 | 0 | 3 |
| 56047 | 0 | 3 |
| 56053 | 0 | 3 |
| 56059 | 4 | 0 |
| 56104 | 0 | 4 |
| 56112 | 55 | 2 |
| 56204 | 0 | 2 |
| 56219 | 4 | 5 |
| 56235 | 4 | 15 |
| 56281 | 3 | 5 |
| 56300 | 6 | 14 |
| 56317 | 0 | 5 |
| 56425 | 4 | 0 |
| 56542 | 2 | 0 |
| 56634 | 0 | 2 |
| 56655 | 0 | 4 |
| 56672 | 0 | 8 |
| 56752 | 2 | 1 |
| 56784 | 10 | 18 |
| 56801 | 2 | 6 |
| 56808 | 2 | 3 |
| 56815 | 0 | 4 |
| 56988 | 0 | 6 |
| 57196 | 1 | 0 |
| 57667 | 19 | 15 |
| 57701 | 12 | 41 |
| 57716 | 4 | 10 |
| 57756 | 2 | 2 |
| 57783 | 0 | 5 |
| 57829 | 8 | 0 |
| 57850 | 4 | 25 |
| 57863 | 2 | 4 |
| 57868 | 0 | 2 |
| 57879 | 2 | 1 |
| 57903 | 2 | 11 |
| 63380 | 2 | 4 |
| 63393 | 0 | 2 |
| 63411 | 0 | 7 |
| 63737 | 2 | 1 |
| 63748 | 0 | 3 |
| 64005 | 0 | 3 |
| 64153 | 0 | 2 |
| 64197 | 2 | 0 |
| 64268 | 1 | 9 |
| 64356 | 0 | 12 |
| 64484 | 0 | 6 |
| 68906 | 2 | 2 |
| 69000 | 0 | 2 |
| 69322 | 1 | 0 |
| 69572 | 2 | 7 |
| 69593 | 2 | 3 |
| 69645 | 0 | 2 |
| 74650 | 0 | 3 |
| 74924 | 0 | 3 |
| 75217 | 2 | 1 |
| 75304 | 0 | 2 |
| 75553 | 0 | 4 |
| 75620 | 0 | 2 |
| 75632 | 1 | 0 |
| 81271 | 0 | 2 |
| 81296 | 6 | 2 |
| 81321 | 2 | 1 |
| 81333 | 0 | 4 |
| 81338 | 0 | 5 |
| 81342 | 0 | 2 |
| 81343 | 2 | 1 |
| 81383 | 0 | 4 |
| 81406 | 2 | 1 |
| 81442 | 0 | 4 |
| 81606 | 2 | 5 |
| 81631 | 2 | 0 |
| 81653 | 2 | 2 |
| 81660 | 2 | 2 |
| 81665 | 0 | 2 |
| 81811 | 2 | 0 |
| 81901 | 2 | 3 |
| 81995 | 2 | 0 |
| 82061 | 1 | 0 |
| 82109 | 2 | 5 |
| 82199 | 2 | 9 |
| 82215 | 2 | 1 |
| 82282 | 2 | 1 |
| 82549 | 0 | 2 |
| 82649 | 2 | 0 |
| 82657 | 0 | 3 |
| 82754 | 2 | 0 |
| 82919 | 2 | 0 |
| 88697 | 1 | 2 |
| 88713 | 0 | 2 |
| 88723 | 0 | 6 |
| 88798 | 0 | 2 |
| 89000 | 4 | 0 |
| 89009 | 4 | 1 |
| 89124 | 0 | 3 |
| 89360 | 0 | 2 |
| 89942 | 0 | 2 |
| 90096 | 0 | 2 |
| 90123 | 2 | 0 |
| 90229 | 2 | 0 |
| 100920 | 6 | 1 |
| 100934 | 2 | 3 |
| 100950 | 4 | 6 |
| 101152 | 0 | 2 |
| 101447 | 0 | 3 |
| 101473 | 0 | 2 |
| 101489 | 2 | 0 |
| 101576 | 2 | 1 |
| 102114 | 0 | 2 |
| 102712 | 0 | 2 |
| 102860 | 2 | 0 |
| 103111 | 0 | 2 |
| 103243 | 2 | 0 |
| 103473 | 22 | 68 |
| 103481 | 3 | 3 |
| 103506 | 10 | 24 |
| 109993 | 0 | 2 |
| 110296 | 0 | 6 |
| 110921 | 2 | 0 |
| 110942 | 2 | 0 |
| 111331 | 2 | 1 |
| 111352 | 2 | 1 |
| 111430 | 2 | 1 |

Table S3: **List of the SCOP superfamily domains appearing in duplications and their relative population in the WGD and non-WGD sets of duplicates.**